\documentclass[aps, prl, reprint, superscriptaddress]{revtex4-2}

\usepackage{amsmath}
\usepackage{amssymb}
\usepackage{bm}
\usepackage{physics}
\usepackage{verbatim}

\usepackage[scr]{rsfso}

\usepackage{graphicx}
\usepackage{overpic}

\usepackage[dvipsnames]{xcolor}
\definecolor{urlblue}{RGB}{6,69,173}

\usepackage[
	colorlinks=true,
	linkcolor=blue,
	anchorcolor=blue,
	citecolor=blue,
	urlcolor=blue,
	pdftitle={latticegraviton!}
]{hyperref}

\newcommand{\ii}{\mathrm{i}}
\newcommand{\ee}{\mathrm{e}}
\newcommand{\hc}{\mathrm{h.c.}\, }
\newcommand{\rr}{\mathbf{r}}
\newcommand{\ex}{\mathbf{e}_{x}}
\newcommand{\ey}{\mathbf{e}_{y}}

\newcommand{\mbf}[1]{\mathbf{#1}}
\newcommand{\mbb}[1]{\mathbb{#1}}
\newcommand{\mc}[1]{\mathcal{#1}}
	
\newcommand{\mrm}[1]{\mathrm{#1}}
\newcommand{\mscr}[1]{\mathscr{#1}}

\renewcommand{\paragraph}[1]{\textit{#1}.---}

\makeatletter
\newcommand{\thetitle}{\@title}
\makeatother

\begin{document}

\title{Chiral Gravitons on the Lattice}
\author{Hernan B. Xavier}
\affiliation{ICTP --- The Abdus Salam International Centre for Theoretical Physics, Strada Costiera 11, 34151 Trieste, Italy}
\affiliation{SISSA --- International School for Advances Studies, via Bonomea 265, 34136 Trieste, Italy}

\author{Zeno Bacciconi}
\affiliation{SISSA --- International School for Advances Studies, via Bonomea 265, 34136 Trieste, Italy}
\affiliation{ICTP --- The Abdus Salam International Centre for Theoretical Physics, Strada Costiera 11, 34151 Trieste, Italy}

\author{Titas Chanda}
\affiliation{Department of Physics, Indian Institute of Technology Madras, Chennai 600036, India}
\affiliation{Center for Quantum Information, Communication and Computation (CQuICC),
Indian Institute of Technology Madras, Chennai 600036, India}

\author{Dam Thanh Son}
\affiliation{Kadanoff Center for Theoretical Physics, University of Chicago, Chicago, Illinois 60637, USA}

\author{Marcello Dalmonte}
\affiliation{ICTP --- The Abdus Salam International Centre for Theoretical Physics, Strada Costiera 11, 34151 Trieste, Italy}

\date{\today}

\begin{abstract}

Chiral graviton modes are elusive excitations arising from the hidden quantum geometry of fractional quantum Hall states. It remains unclear, however, whether this picture extends to lattice models, where continuum translations are broken and additional quasiparticle decay channels arise. We present a framework in which we explicitly derive a field theory incorporating lattice chiral graviton operators within the paradigmatic bosonic Harper-Hofstadter model. Extensive numerical evidence suggests that chiral graviton modes persist away from the continuum, and are well captured by the proposed lattice operators. We identify geometric quenches as a viable experimental probe, paving the way for the exploration of chiral gravitons in near-term quantum simulation experiments.

\end{abstract}
\maketitle
\paragraph{Introduction}There is currently considerable interest in exploring topological phenomena in synthetic quantum platforms \cite{cooper2019topological,ozawa2019topological,banuls2020simulating,zhang2018topological}, which offer fundamental and complementary information in terms of probes and modeling, compared to widely successful solid state experiments \cite{haldane2017nobel,fradkin2013field,moessner2021topological,zeng2019quantum}.
In particular, cold atoms in optical potentials have shown promising developments in realizing lattice models pierced by synthetic magnetic fields, broadly belonging to the dynamics of the paradigmatic Harper-Hofstadter Hamiltonian \cite{hofstadter1976energy,jaksch2003creation}.
The exploration of the fractional quantum Hall effect (FQHE), and, more broadly, strongly interacting topological matter, remains an overarching goal of such endeavor \cite{gerster2017fractional,motruk2017phase,rosson2018bosonic,schoonderwoerd2019interaction,palm2021pfaffian,boesl_prb2022_topomott,zerba2024emergent,nardin2024quantum,lunt_prl2024_fqhfermions}.
Building on the theory proposal of utilizing laser fields to modify atomic motions \cite{jaksch2003creation}, early experiments have observed key properties of Harper-Hofstadter models \cite{aidelsburger2013realization,miyake2013realizing}, from measuring the Chern number of its bands \cite{neupert2011fractional,scaffidi2014exact,aidelsburger2015measuring}, to demonstrating and visualizing chiral edge currents in both fermionic \cite{mancini2015observation} and bosonic \cite{stuhl2015visualizing} ribbons, as well as the chiral motion of interacting excitations \cite{lienhard2020realization}. More recently, a remarkable study \cite{impertro2024realization} demonstrated coherent dynamics of interacting particles, a feat that remained so far elusive due to the intrinsic difficulty of combining interactions with time-dependent fields needed to introduce an effective magnetic field. 

These recent developments naturally raise the question: to what extent can future experiments provide insight into genuine features of the FQHE? 
One particularly intriguing direction is the possibility of probing its rich spectral structure, including the presence of the recently proposed \cite{haldane2009hall,haldane2011geometrical,golkar2016spectral, son2019chiral,yang2016acoustic} and observed \cite{liang2024evidence} chiral graviton mode excitations. 
Chiral gravitons are spin-2 excitations stemming from the quantum geometric nature of quantum Hall wave functions \cite{laughlin1983anomalous}, which, despite being usually embedded into the continuum \cite{gmp1986magnetoroton}, 
seems to remain a well defined quasiparticle whose decay into \textit{dipolar} excitations is suppressed \cite{golkar2016spectral,gromov2017bimetric}.
Whether such excitations persist in lattice models of the FQHE -- and if so, \textit{how} they can be probed -- remains an open question.
A central challenge lies in identifying a low-energy emergent symmetry on the lattice \cite{du2022volume,wang2025dynamics}, that could protect gravitons from inevitable decay into the continuum due to the breaking of translational invariance induced by the lattice.

\begin{figure}
    \centering
    \begin{overpic}[width=.99\linewidth]{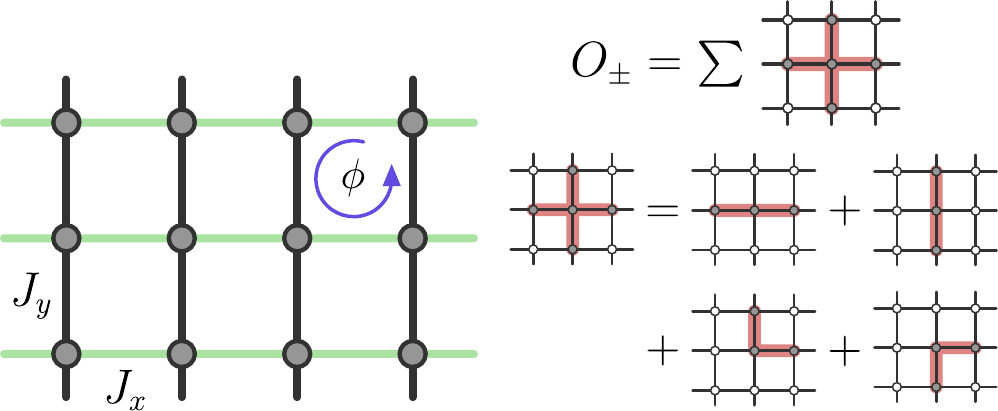}
        \put(-1,34){(a)}
        \put(52,6){(b)}
    \end{overpic}
    \caption{Schematic representations of (a) the Harper-Hofstadter model and (b) the graviton operators we propose. The lattice model has tunneling rates $J_x$ and $J_y$ along perpendicular directions, and magnetic flux $\phi$ per plaquette. Chiral graviton operators are expressed as the sum of different correlated hopping terms centered at a given site.}
    \label{fig:model}
\end{figure}

In this work, we report the existence of an emergent graviton mode in bosonic $\nu=1/2$ Harper-Hofstadter models at diverse fluxes. 
We start by formulating a lattice ansatz for the chiral graviton operators, distilling fundamental ingredients out of its continuum counterpart: the chiral components of the kinetic stress tensor within the lowest Landau level (LLL) \cite{nguyen2014lowest,nguyen2021probing}.
We systematically corroborate our ansatz through an extensive numerical analysis, which combines exact diagonalization (ED) and matrix product state (MPS)~\cite{Schollwock2011, Orus2014} methods. 
We find clear signatures for the graviton mode by computing the spectral densities of the proposed lattice operators in various geometries.
Finally, we discuss concrete protocols for the detection of gravitons in cold atom arrays. In particular, we describe a \textit{geometric quench} \cite{ippoliti2018geometry,liu2018quench,liu2021quench}, that activates the graviton mode highlighting the connection of this mode to the quadrupolar charge correlations in the fluid \cite{haldane2011geometrical,long2024fci}. Additional information and method details are made available in the Supplemental Material (SM) \cite{suppmat}.

\paragraph{Lattice model}We study the bosonic Harper-Hofstadter model, with generic anisotropic tunneling rates $J_x=J$ and $J_y=\eta J$ along $x$ and $y$. The Hamiltonian is written as
\begin{equation}\label{eq:model}
    H=-J\sum_{\rr}\Big(
    b_{\rr}^{\dagger}b_{\rr+\ex}\ee^{\ii2\pi n_\phi y/a}
    +\eta b_{\rr}^{\dagger}b_{\rr+\ey}
    +\hc\Big),
\end{equation}
where $b_\rr^{\dagger}$ and $b_\rr$ are the creation and annihilation operators at site $\rr$, $n_\phi$ is the flux number, and $a$ the unit spacing. We incorporate interactions by implementing the hardcore constraint: $b_\rr^2=0$. We consider rational fluxes of the form $n_\phi=1/q$ (with $q\in\mbb{N}^+$) and magnetic filling $\nu=1/2$, achieved by setting the density of particles to $\rho_0= \nu n_\phi$. This model is readily accessible in cold-atom \cite{miyake2013realizing, leonard2023realization} and superconducting qubits \cite{roushan2017chiral,rosen2024synthetic} platforms.

\paragraph{The continuum limit}We establish a link to the continuum theory of the FQHE. The continuum limit correspond to the case where $a/\ell_B= \sqrt{2\pi n_\phi}$ goes to zero. To take into account interactions coming from the hard-core constraint on the lattice, we employ a Holstein-Primakoff transformation \cite{fradkin2013field}, and
motivated by the dilute limit of the system in the low-flux limit $\rho_0\to0$, we expand in powers of the local density $n_\rr=b_\rr^{\dagger}b_\rr$.
With this, we write:
\begin{equation}\label{eq:mode-expansion}
    \frac{b_\rr}{a}\approx
    \psi(\rr)
    -\frac{a^2}{2}\psi^\dagger(\rr)\psi(\rr)\psi(\rr)
    +\cdots,
\end{equation}
where $\psi(\rr)$ is a continuum bosonic field capturing collective (spin wave) excitations on top the reference vacuum state on the lattice. Replacing this expression into the lattice Hamiltonian of Eq.~(\ref{eq:model}), 
we arrive at the continuum model:
\begin{equation}\label{eq:Hcont}
    H=\int\mathrm{d}^2r\Big(-\frac{g^{ij}}{2m}\psi^\dagger D_i D_j\psi
    +\frac{\kappa}{2}\psi^\dagger\psi^\dagger\psi\psi\Big)
    +\cdots,
\end{equation}
where we introduce the covariant derivatives $D_x=\partial_x+\ii y/\ell_B^2$ and $D_y=\partial_y$, as well as the unimodular rank-2 tensor $g_{ij}$, with nonzero components $g_{xx}=1/g_{yy}=\sqrt{\eta}$.
The effective band mass reads 
$m=1/2\sqrt{\eta}Ja^2$, 
while the strength of interactions scales as 
$\kappa=4(1+\eta)Ja^2$.
As a result, the cyclotron frequency is
$\omega_B=1/m\ell_B^2=4\pi\sqrt{\eta}Jn_\phi$,
and the effective isotropic pseudopotential interaction within the LLL has strength 
$V_0=\kappa/2\pi\ell_B^2=4(1+\eta)Jn_\phi$ \cite{yang2017generalized}. 
The continuum model has two Ward identities we exploit below. They concern charge $\rho=|\psi|^2$ and momentum conservation:
\begin{equation}\label{eq:wardidentities}
    \partial_t\rho+\nabla\cdot\bm{j}=0,\qquad
    \partial_t(m\bm{j})+\nabla\cdot\overset{\leftrightarrow}{\bm{\mc{T}}}=\bm{f}_\mrm{ext},
\end{equation}
where $\bm{j}$ is the particle current, and $\overset{\leftrightarrow}{\bm{\mc{T}}}$ is the stress tensor, whose divergence bounds the flux of momentum, modulo external forces $\bm{f}_\mrm{ext}$.

Proceeding to a LLL description, graviton operators are associated with the chiral components of the stress tensor \cite{liou2019chiral,nguyen2014lowest}.
The LLL limit is equivalent to a zero mass approximation $m\to0$ \cite{du2022volume}, where the free kinetic term gets quenched.
To simplify matters we consider the isotropic limit ($\eta\to1$).
In this case, the continuum, long-distance field theory is written as follows:
\begin{align}
    \mscr{L}
    &=\psi^\dagger\ii\partial_t\psi
    +2\ell_B^2\chi^\dagger D_{z}D_{\bar{z}}\psi
    \nonumber\\
    &
    +2\ell_B^2D_{\bar{z}}D_{z}\psi^\dagger\chi
    -\frac{\kappa}{2}|\psi|^4.
\end{align}
The LLL projection is enforced by the introduction of the Lagrange multiplier $\chi$ \cite{nguyen2021probing}, which ensures that $\psi$ is a linear superposition of wavefunctions of the form $\ee^{\ii k_xx -(y+k_x\ell_B^2)/2\ell_B^2}$.
The complex variables are defined as $z=x+\ii y$ and $\bar{z}=x-\ii y$, with the corresponding holomorphic derivatives are given by 
$D_{z}=\partial_z-\ii A_z$ and $D_{\bar{z}}=\partial_{\bar{z}}-\ii A_{\bar{z}}$, where $A_{z}=A_{\bar{z}}=-y/2\ell_B^2$ in the cylindrical gauge. The traceless part of the stress tensor is obtained from Noether’s theorem:
\begin{equation}\label{eq:Tlagrange}
    \mc{T}_{zz}=-\ell_B^2\chi^\dagger D_{z}^2\psi,\qquad
    \mc{T}_{\bar{z}\bar{z}}=-\ell_B^2 D_{\bar{z}}^2\psi^\dagger\chi,
\end{equation}
where we use the fact that the operator $D_{\bar{z}}$ annihilates $\psi$ to simplify our formulas. To eliminate the Lagrange multiplier we minimize the action with respect to $\psi^\dagger$, which leads us to the Schrödinger-type equation:
\begin{equation}
    \ii\partial_t\psi
    +2\ell_B^2 D_{z}D_{\bar{z}}\chi
    -\kappa|\psi|^2\psi=0.
\end{equation}
To solve the constraint, one must expand field operators onto different Landau levels \cite{nguyen2021probing}, e.g., $\psi=\psi_0+\psi_1+\cdots$, and as we are interested in the LLL theory, then solve the equation for $\chi_2=-\kappa|\psi|^2\psi/2$ that enters Eq.~(\ref{eq:Tlagrange}). In this form, we arrive at the chiral components of the stress tensor:
\begin{align}
    \mc{T}_{zz}
    &=\frac{\kappa\ell_B^2}{2}\psi^{\dagger}\psi^{\dagger}\psi(D_{z}^2\psi),
    \nonumber\\
    \mc{T}_{\bar{z}\bar{z}}
    &=\frac{\kappa\ell_B^2}{2}(D_{\bar{z}}^2\psi^{\dagger})\psi^{\dagger}\psi\psi.
\end{align}
The chiral graviton operators derive from the long wavelength mode of these operators, with $\mc{O}_-=\mc{T}_{zz}(\mbf{q}\to0)$ and $\mc{O}_+=\mc{T}_{\bar{z}\bar{z}}(\mbf{q}\to0)$. In momentum space, these operators assume the form \cite{liou2019chiral}:
\begin{equation}
    \mc{O}_{\pm}\propto
    \sum_{\mbf{q}}(q_x\pm\ii q_y)^2\ee^{-q^2\ell_B^2/2}\bar{\rho}_{\mbf{q}}\bar{\rho}_{-\mbf{q}},
\end{equation}
where $\rho_\mbf{q}$ is the LLL guiding-center density operator.

\paragraph{Lattice stress tensor}We now return to the isotropic lattice model to look for operator representations of $\mc{T}_{zz}$ and $\mc{T}_{\bar{z}\bar{z}}$. The strategy is to inspect the momentum flux in the lattice, from where we bootstrap the operator that matches the stress tensor in the continuum. 
Given charge conservation is also a symmetry of the lattice model, first we match the current operators. Using the discrete continuity equation, we obtain:
\begin{align}\label{eq:lattice-current}
j_x(\rr)
&=
-\frac{\ii J}{a}\big(
b_{\rr}^{\dagger}b_{\rr+\ex} \ee^{\ii2\pi n_\phi y/a}
-\hc
\big),
\nonumber\\
j_y(\rr)
&=
-\frac{\ii J}{a}\big(
b_{\rr}^{\dagger}b_{\rr+\ey} 
-\hc
\big),
\end{align}
where we add an $1/a$ factor that reflects its engineering dimension. Performing the low-density expansion, we find the current operators correctly reproduce: $\mbf{j}(\rr)=\bm{j}(\rr)+\cdots$, where
$\bm{j}(\rr)=\frac{1}{2m}\big[\psi^\dagger(-\ii\overset{\leftrightarrow}{\nabla})\psi-2e\mbf{A}|\psi|^2\big]$, and $A\overset{\leftrightarrow}{\nabla}B=A(\nabla B)-(\nabla A)B$, and all other parameters follow the definitions given around Eq. (\ref{eq:Hcont}).

The second Ward identity, Eq. (\ref{eq:wardidentities}), only emerges in the continuum. Two key points are 
(i) this is not related to a conservation law in the lattice;
and (ii) any direct computation of $\partial_tj_i=\ii[H,j_i]$ shall produce not only differences of $T_{ij}$ but also an external force contribution.
To circumvent these limitations we combine long-distance properties with a lattice trick. 
First we consider diagonal terms $T_{xx}$ and $T_{yy}$, which are not affected by an external force of Lorentz type $f_i\propto \epsilon_{ij}j_j$.
We isolate this contribution by splitting the Hamiltonian into terms that generate motion along $x$ and $y$: $H=H_x+H_y$. In this way the $T_{ii}$ component can be directly identified from the evaluation of $\Delta T_{ii}=-\ii m\comm{H_i}{j_i}$. 
Note that the expressions depend explicitly on the effective mass $m$, as we use continuum symmetry considerations to guide our lattice calculations. Taking the $xx$-component as an example, from this we find:
\begin{align}
    T_{xx}^{full}(\rr)
    &=
    mJ^2
    \Big[
    b_{\rr-\ex}^{\dagger}(2n_{\rr}-1)b_{\rr+\ex}\ee^{\ii4\pi n_\phi y/a}
    +\hc
    \nonumber\\
    &
    +2n_\rr\Big].
\end{align}
We call this the full stress tensor as it carries a free contribution, quenched in the LLL projection. We can get rid of it by repeating the calculation with free bosons, and taking the difference. By doing this, we kill all but the correlated hopping terms:
\begin{equation}
T_{xx}(\rr)
=
J
\Big(
b_{\rr-\ex}^{\dagger}n_{\rr}b_{\rr+\ex}\ee^{\ii4\pi n_\phi y/a}
+\hc
\Big),
\end{equation}
where we substitute $m=1/2Ja^2$, and set the lattice spacing to unity.
An analogous computation for the $yy$-component leads us to the identification of
\begin{equation}
T_{yy}(\rr)
=
J
\Big(
b_{\rr-\ey}^{\dagger}n_{\rr}b_{\rr+\ey}
+\hc
\Big).
\end{equation}
In principle, similar arguments should be applicable to the computation of the off-diagonal component $T_{xy}$. We instead construct $T_{xy}$ from the insight we just developed. In analogy to the diagonal components, the off-diagonal $T_{xy}$ must arise from diverting the current by generating jumps across its perpendicular direction, what motivates L-shaped correlated hopping moves, see Fig.~\ref{fig:model}. Broken parity implies this component should be odd under this symmetry transformation. Putting all together, we propose: 
\begin{align}
    T_{xy}(\rr)
    &=
    J\Big(
    b_{\rr-\ey}^{\dagger}n_\rr b_{\rr+\ex}\ee^{\ii2\pi n_\phi y/a}
    \nonumber\\
    &-b_{\rr+\ey}^{\dagger}n_\rr b_{\rr+\ex}\ee^{\ii2\pi n_\phi y/a}
    +\hc
    \Big).
\end{align}
The overall constant can be checked via the low-density, continuum expansion. For instance, for the $zz$-component we indeed find:
$T_{zz}\sim\mc{T}_{zz}$. We then propose lattice forms for the chiral graviton operators $O_+$ and $O_-$, constructing them from the long-wavelength components of the stress tensor:
\begin{equation}\label{eq:grav-operators}
O_{\pm}=
\sum_{\rr}
\Big[T_{xx}(\rr)-T_{yy}(\rr)\pm2\ii T_{xy}(\rr)\Big].
\end{equation}

\begin{figure}
    \centering
    \begin{overpic}[width=\linewidth]{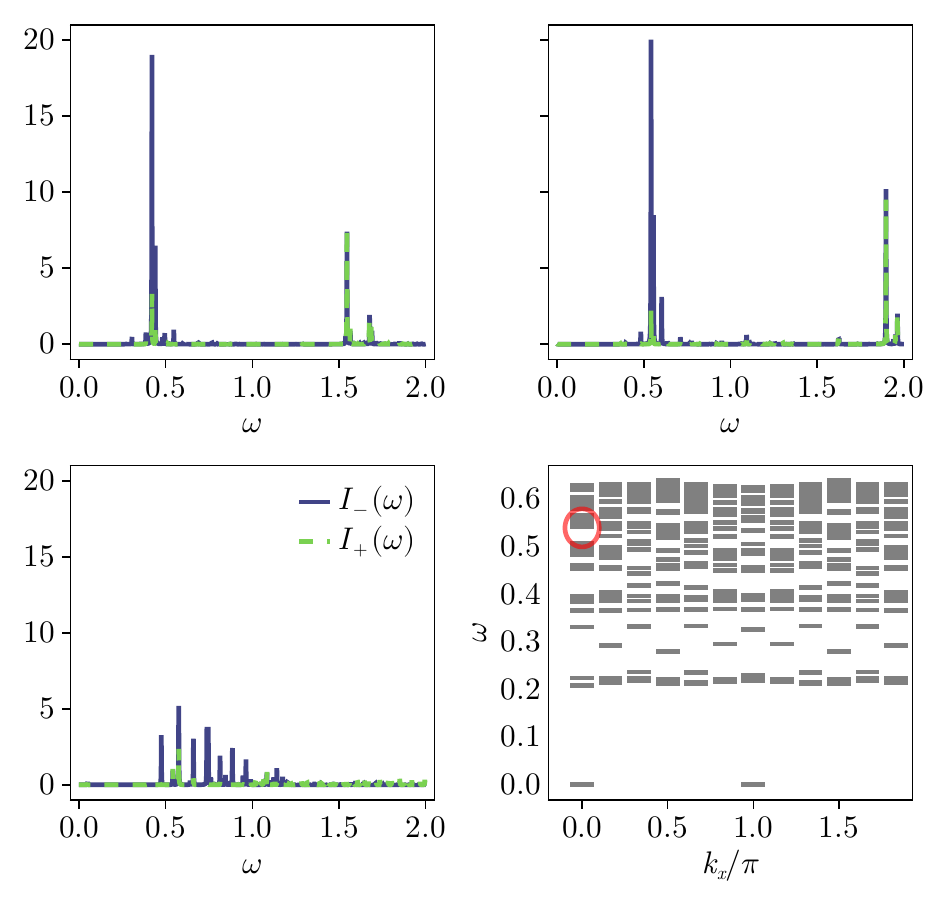}
        \put(37,88){(a)}
        \put(30,80){\footnotesize$n_\phi=1/8$}
        \put(30,75){\footnotesize$12\times8$}
        \put(88,88){(b)}
        \put(81,80){\footnotesize$n_\phi=1/6$}
        \put(81,75){\footnotesize$12\times6$}
        \put(11,40){(c)}
        \put(30,30){\footnotesize$n_\phi=1/4$}
        \put(30,25){\footnotesize$8\times8$}
        \put(88,16){(d)}
    \end{overpic}
    \caption{Graviton weights $I_\pm(\omega)$ for:
    (a) $12\times8$ torus with $n_\phi=1/8$ flux;
    (b) $12\times6$ torus with $n_\phi=1/6$ flux; and
    (c) $8\times8$ torus with $n_\phi=1/4$ flux. 
    (d) Low-energy spectrum for a $12\times6$ torus. The red circle indicates the energy scale of the graviton-mode, $\omega_\mrm{G}\approx0.54$.
    All results extracted from ED methods. Spectral densities obtained by taking $n_\mrm{Krylov}=1000$ Krylov states, and a Lorentzian broadening factor of $\gamma=0.001$.
    }
    \label{fig:weights}
\end{figure}
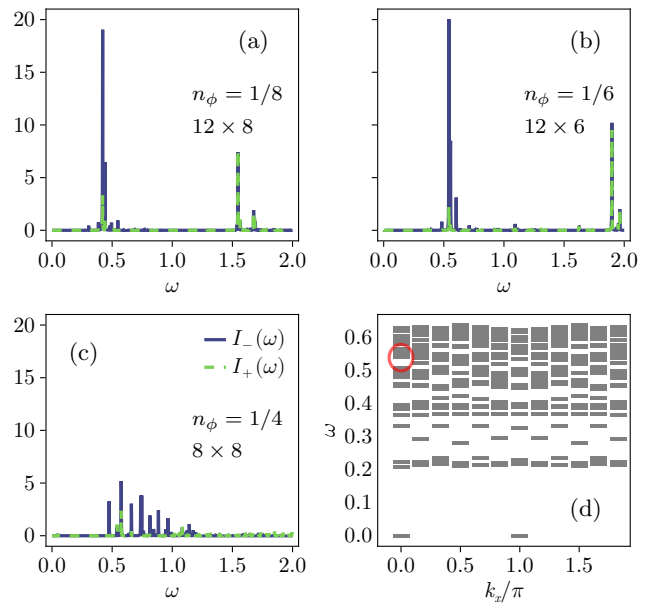

\paragraph{Graviton weights}We now numerically inspect the spectral density of the lattice chiral graviton operators in Eq. (\ref{eq:grav-operators}) at the isotropic point ($\eta=1$). Their weights are defined as:
\begin{equation}\label{eq:grav_weights}
I_\pm(\omega)=\frac{1}{\mc{N}}\sum_{n}|\bra{n}O_{\pm}\ket{0}|^2\delta(\omega-E_n),
\end{equation}
where $E_n$ represents the excitation energy of the many-body state $\ket{n}$, and the normalization constant ensures $\int_{-\infty}^{+\infty}\dd\omega\, I_\pm(\omega)=1$. In Fig. \ref{fig:weights} we show ED results on the chiral graviton weights for different fluxes (a) $n_\phi=1/8$, (b) $n_\phi=1/6$, and (c) $n_\phi=1/4$. All results are obtained on the torus geometry at commensurate filling \footnote{Commensurate filling is obtained by loading the lattice with $N=\nu\times n_\phi\times A$ particles, with $A=L_x\times L_y$ the number of sites.}. 
Coming from low to high fluxes, we first consider the cases with $n_\phi=1/8$ and $n_\phi=1/6$. We observe a great resemblance in both scenarios, which feature a strong chiral signal. At low frequencies, most of the $I_-(\omega)$ weights are accumulated by a few peaks around $\omega\approx0.5J$. We highlight the nontrivial character of such an emergent, spin-2 excitation in a lattice model which only counts with $C_2$, 180 degrees rotations. These states are hidden into the continuum, as shown explicitly in Fig. \ref{fig:weights}(d) where we plot the many-body spectrum for $n_\phi=1/6$. The graviton-mode energy $\omega_\mrm{G}$ is slightly more than double the many-body gap, set by magnetoroton excitations $\omega_\mrm{R}$ \cite{liou2019chiral, kumar2022neutral,liu2024resolving}. 
Both $\omega_\mrm{R}$ and $\omega_\mrm{G}$ shift up as the flux $n_\phi$ is increased (see SM for more details), agreeing with the continuum limit effective model prediction, which sets the energy scale of interactions to $V_0=8Jn_\phi$. 
A rough linear fit of these values indicates the energy scale of the graviton-mode is given by $\omega_\mrm{G}\approx0.42\times V_0$ \cite{liu2018quench}.
A second, nonchiral response is found at higher energies which roughly match the single-particle gap dictated by the continuum cyclotron frequency $\omega_B=4\pi Jn_\phi$. 

The graviton spectrum seems to drastically change once the flux reaches $n_\phi=1/4$, as shown in Fig. \ref{fig:weights}(c). Even though the response remains chiral, the weight is now distributed on the whole continuum of states instead of a showing a sharp response as for the lower flux cases. We also find a similar chiral but broad graviton response function at $n_\phi=1/4$ computed via the time-dependent variational principle (TDVP)~\cite{Haegeman2011, Haegeman2016, Paeckel2019} on long open cylinders \cite{suppmat}. 
It however remains not clear whether the mixing of the graviton mode with the many-body continuum at $n_\phi=1/4$ is a fundamental feature of the lattice model, or a finite size effect.

\begin{figure}
    \centering
    \begin{overpic}[width=\linewidth]{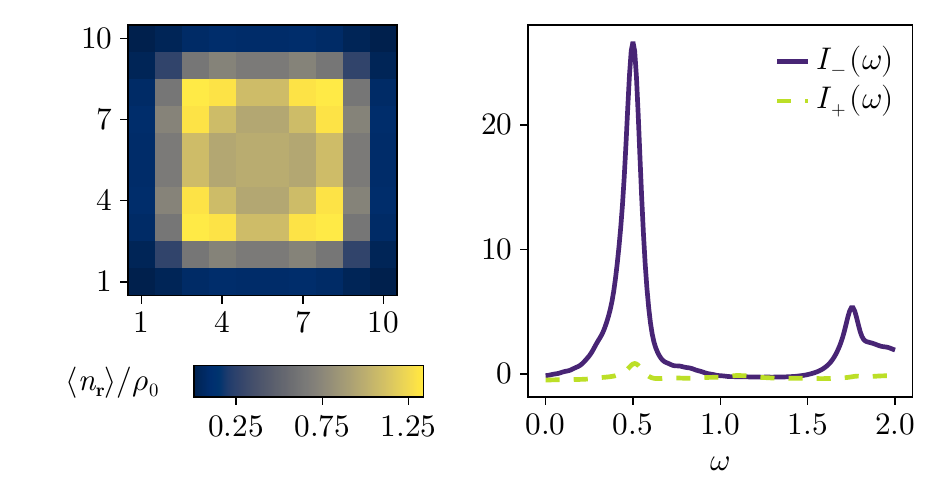}
        \put(1,14){(a)}
        \put(74,36){(b)}
    \end{overpic}
    \caption{Graviton mode on a FQHE droplet for $n_\phi=1/6$ with $N=5$ particles. 
    (a) Expectation value of the density in the ground state, $\ev{n_\mbf{r}}$. 
    (b) Graviton response obtained through time evolution with a Lorentzian broadening of $\gamma=0.05$.
    Total duration of $T=100/J$, performed in time steps of $\delta t=0.1/J$.
    MPS results obtained with a maximum bond dimension $\chi_\mrm{max}=300$, attaining truncation errors of the order $\varepsilon_\mrm{trunc}\sim10^{-6}$.}
    \label{fig:droplet}
\end{figure}

\paragraph{Open edges}Next we address the resilience of the chiral graviton response on a small FQHE droplet, relevant for current quantum simulation platforms \cite{leonard2023realization,impertro2024realization}. 
We use MPS methods to investigate a plane of size $10\times10$, with edges left open. The flux is that of $n_\phi=1/6$ per plaquette, and we load the lattice with $N=5$ particles \cite{suppmat}. 
Figure \ref{fig:droplet}(a) shows the (normalized) local density on the ground state obtained with density matrix renormalization group (DMRG)~\cite{White1992, White1993} simulations. 
Density oscillations sensitive to the system's geometry arise close to the hard egdes, but the bulk displays a tendency towards the uniform density $\ev{n_\mbf{r}}/\rho_0\approx1$.
The spectral weights are shown in Fig. \ref{fig:droplet}(b), which are obtained via TDVP \cite{suppmat}. Note that in open geometries $O_\pm$ acquire finite expectation values which must be subtracted according to the definition of the graviton weights in Eq.~(\ref{eq:grav_weights}).
We observe a strong graviton signal, comparable to that found in the torus geometry of Fig.~\ref{fig:weights}(c), indicating that such excitations remain robust and detectable even in small lattice droplets.

\begin{figure*}
    \centering
    \begin{overpic}[width=0.98\linewidth]{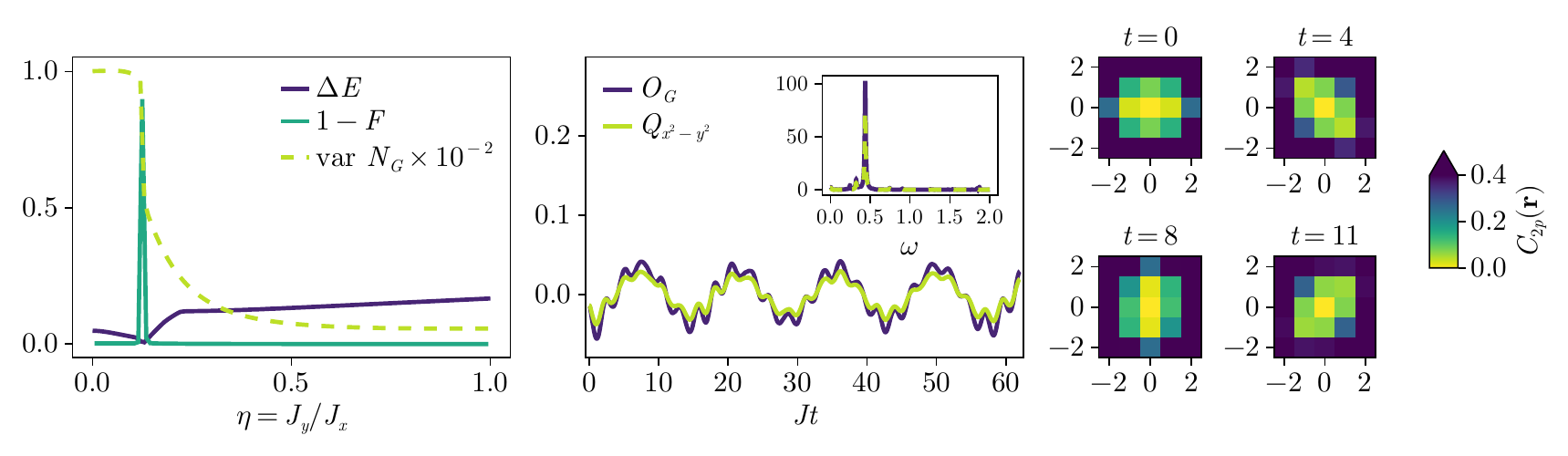}
        \put(27,13){(a)}
        \put(40,16){(b)}
        \put(90,24){(c)}
    \end{overpic}
    \caption{Anisotropy study for $n_\phi=1/8$ flux. 
    (a) Ground state phase diagram characterization. Plot shows the behavior of energy gap, fidelity, and variance of the graviton number as a function of the anisotropy $\eta$ in the $k_x=0$ momentum sector.
    ED results for a $10\times8$ torus.
    (b) Quench dynamics on a $8\times8$ torus. Real time behavior of the monitored observables. The inset shows the Fourier transform obtained after a time evolution $T=600/J$, with a damping $\gamma=5/T$.
    (c) Snapshots of the two-particle function $C_{2p}(\mbf{r})$ showing the clockwise precession of the correlation hole during the quench.}
    \label{fig:quench}
\end{figure*}

\paragraph{Quench dynamics}Geometric quenches, i.e., sudden changes in the mass anisotropy $g^{ij}$, have been proposed in continuum models as a simple way to activate graviton-modes \cite{ippoliti2018geometry,liu2018quench,liu2021quench}. This can be readily generalized to the lattice case via sudden changes of the tunneling ratios $\eta=J_y/J_x$, which are directly linked to $g^{ij}$ in the continuum limit, see Eq. (\ref{eq:Hcont}). As opposed to solid-state systems, such protocols are very natural for cold-atom experiments and quantum simulators in general.
We first test the endurance of the FQHE phase on the lattice as $\eta$ is varied. In Fig. \ref{fig:quench}(a) we show ED results obtained for a torus of size $10\times8$ and flux $n_\phi=1/8$. They indicate a transition only occurs at small $\eta\approx0.12$, where the excitation gap $\Delta E$ closes and the ground state fidelity $F(\eta)=|\ev{\psi(\eta)|\psi(\eta+\delta\eta)}|^2$ drops.
In the same plot we further include the variance of the number-like operator $N_\mrm{G}=\frac12(O_{-}O_{+}+O_{+}O_{-})$, which not only captures the transition but its lower variance inside the FQHE phase is suggestive of an approximate symmetry leading to a well-defined graviton mode. 
We emphasize these results are compatible with larger volume, MPS simulations on finite cylinders \cite{suppmat}.
Having set up a window of phase stability, we consider a slingshot dynamics around the isotropic point. We first prepare an anisotropic ground state with $\eta=0.6$, and then change suddenly $\eta\to1$. We choose to monitor two observables: the so called graviton displacement $O_\mrm{G}=O_++O_-$ \cite{liou2019chiral}, and the quadrupolar density-density correlations $Q_{x^2-y^2}=\sum_\rr(n_\rr n_{\rr+\ex}-n_\rr n_{\rr+\ey})$ \cite{long2024fci}. 
Figure \ref{fig:quench}(b) shows the real-time dynamics of these observables for early times. We see both have very similar characteristics. Indicative of a close relationship among the graviton mode and the quadrupole density correlations, as occurs in continuum instances of the FQHE.
Collecting data up to the final time $T=600/J$ ultimately leads to the sharp Fourier responses shown in the inset.
The two-particle function 
$C_{2p}(\mbf{r}-\mbf{r}')
=\ev*{b_\mbf{r}^{\dagger}b_{\mbf{r}'}^{\dagger}b_{\mbf{r}'}^{\phantom\dagger}b_\mbf{r}^{\phantom\dagger}}/\rho_0^2$ 
is another informative quantity during the dynamics. In Fig. \ref{fig:quench}(c) we show four snapshots of $C_{2p}(\mbf{r})$ taken at different times. The snapshots reveal the clockwise precession of the elliptical correlation hole, with period $T_\mrm{G}=2\pi/\omega_\mrm{G}\approx14/J$. This is to be contrasted with the counterclockwise direction of the flux thus highlighting the negative angular momentum of the graviton-mode.

\paragraph{Conclusions}In this work, we assessed the existence of the chiral graviton mode in a Harper-Hofstadter model with hardcore bosons; a system of direct experimental interest \cite{leonard2023realization}.
Importantly, we introduced a new lattice operator representation for LLL graviton operators \cite{liou2019chiral,nguyen2021probing}, ensuring consistency with the continuum limit -- an approach that could potentially be applied to recent findings in the context of fractional Chern insulators~\cite{neupert2011fractional,long2024fci,wang2025dynamics}.  
To validate our approach, we performed a series of numerical simulations.
For fluxes as high as $n_\phi=1/5$, we found clear evidence of a sharp chiral graviton mode by computing the weights of the proposed lattice operators. 
We conducted additional simulations to (i) test the resilience of the graviton mode in small FQHE droplets and (ii) extract its dynamical response from a geometric quench. These results show that a direct experimental study of graviton-mode in quantum simulators is possible.
Our study leaves several promising directions open for future research. 
Future questions regard the fate of the graviton at high fluxes as the overlap to Laughlin states decreases \cite{sorensen2005fractional,hafezi2007fractional,hafezi2007characterization}, and the extension to other non-Abelian FQHE states which are also realized in the Harper-Hofstadter model \cite{moore1991nonabelions,palm2021pfaffian}.

\paragraph{Acknowledgments}We thank Ajit Balram, Federico Becca, and Philipp Lunt for discussions, and Iacopo Carusotto for collaboration on related work. 
ED computations have been performed in Python with the aid of QuSpin package \cite{QuSpin2017p1,QuSpin2017p2}.
MPS  simulations employed the ITensors.jl \cite{ITensor2022,ITensor2022codebase} and TenNetLib.jl \cite{TenNetLib} libraries available in Julia. 
M.\,D. was partly supported by the QUANTERA DYNAMITE PCI2022-132919, by the EU-Flagship programme Pasquans2, by the PNRR MUR project PE0000023-NQSTI, the PRIN programme (project CoQuS), and the ERC Consolidator grant WaveNets. H.\,B.\,X. was supported by the MIUR programme FARE (MEPH). 
Z.\ B. thanks IIT Madras for kind hospitality through the Centers of Excellence, QuCenDiEM (Project No.\ SP22231244CPETWOQCDHOC) and CQuICC (Project No.\ SP22231228CPETWOCQIHOC).
T.\ C. acknowledges the support by the Young Faculty Initiation Grant (NFIG) at IIT Madras (Project No.\
RF24250775PHNFIG009162).
The work of D.\ T.\ S. is supported, in part, by the U.S.\ DOE grant
No.\ DE-FG02-13ER41958 and by the Simons Collaboration on Ultra-Quantum Matter, which is a grant from the Simons Foundation (No.\ 651442, DTS).
H.\,B.\,X. and M. D. thank the Pollica Physics Centre for hospitality during the early stages of this work, and Leon Balents, Meng Zi-Yang, and Subir Sachdev for stimulating discussion during the workshop ``Exotic quantum matter from quantum spin liquids to novel field theories.''

\bibliography{graviton}

\clearpage
\onecolumngrid

\begin{center}
    {\large\bfseries 
    Supplementary Material for ``\thetitle''
    } \\[0.5cm]
    {
    Hernan B. Xavier, Zeno Bacciconi, Titas Chanda, Dam Thanh Son, and Marcello Dalmonte}
\end{center}
\setcounter{equation}{0}
\setcounter{figure}{0}
\renewcommand{\theequation}{S\arabic{equation}}
\renewcommand{\thefigure}{S\arabic{figure}}
\renewcommand{\thesection}{\arabic{section}}

\section{Contents}

{\hypersetup{hidelinks}
\begin{itemize}
\item[] \hyperref[app:change]{Change of gauge and open geometry simulations} \hfill \pageref{app:change}
\item[] \hyperref[app:more]{More on graviton weights} \hfill \pageref{app:more}
\item[] \hyperref[app:numerical]{Numerical procedure for spectral functions with MPS} \hfill \pageref{app:numerical}
\item[] \hyperref[app:large]{Large volume graviton spectral functions} \hfill \pageref{app:large}
\item[] \hyperref[app:small]{Small volume graviton spectral functions} \hfill \pageref{app:small}
\item[] \hyperref[app:phase]{Phase diagram on cylinders with MPS} \hfill \pageref{app:phase}
\end{itemize}
}

\section{Change of gauge and open geometry simulations}
\label{app:change}

In the main matter we adopt a Landau-type gauge, with a magnetic unit cell that wraps the system along the $y$-direction.
In this supplement, we account for different gauge choices, writing the corresponding formulas for the graviton operators. For reference, consider the isotropic Harper-Hofstadter model:
\begin{equation}
    H=-J\sum_{\rr}
    \Big(
    b_{\rr}^{\dagger}b_{\rr+\ex}\ee^{\ii\theta_{x}y}
    +b_{\rr}^{\dagger}b_{\rr+\ey}\ee^{-\ii\theta_{y}x}
    +\hc
    \Big),
\end{equation}
where $\theta_{x}$ and $\theta_{y}$ are hopping phases as coming from a Peierls substitution.
The counterclockwise circulation around every plaquette bounds the flux:
\begin{equation}
    \theta_{x}+\theta_{y}=2\pi n_\phi.
\end{equation}
We set the lattice spacing to unit $a=1$.
Cylindrical gauge choices are implemented by choosing either $\theta_x$ or $\theta_y$ equal to zero. The symmetric gauge on the other hand corresponds to take $\theta_{x}=\theta_{y}=\pi n_\phi$. All these cases can be summarized into the formulas below. For the diagonal components:
\begin{align}
    T_{xx}(\rr)-T_{yy}(\rr)
    &=J\Big(
        b_{\rr-\ex}^{\dagger}n_\rr b_{\rr+\ex}\ee^{2\ii\theta_{x}y}
        -b_{\rr-\ey}^{\dagger}n_\rr b_{\rr+\ey}\ee^{-2\ii\theta_{y}x}
        +\hc
    \Big),
\end{align}
and for the off-diagonal terms:
\begin{align}
    T_{xy}(\rr)
    &=J\Big(
        b_{\rr-\ey}^{\dagger}n_\rr b_{\rr+\ex}\ee^{\ii\theta_{x} y-\ii\theta_{y}x}
        -b_{\rr+\ey}^{\dagger}n_\rr b_{\rr+\ex}\ee^{\ii\theta_{x} y+\ii\theta_{y}x}
        +\hc
    \Big).
\end{align}
To mitigate the loss of translation invariance in open geometries (see MPS simulations below), we employ a \textit{symmetrized} version of the $T_{xy}$ operator, defined as
\begin{align}
    T_{xy}(\rr)
    &=\frac{J}{2}\Big(
        b_{\rr-\ey}^{\dagger}n_\rr b_{\rr+\ex}\ee^{\ii\theta_{x} y-\ii\theta_{y}x}
        -b_{\rr+\ey}^{\dagger}n_\rr b_{\rr+\ex}\ee^{\ii\theta_{x} y+\ii\theta_{y}x}
        +\hc
    \Big)
    \nonumber\\
    &+\frac{J}{2}\Big(
        b_{\rr-\ex}^{\dagger}n_\rr b_{\rr+\ey}
        \ee^{\ii\theta_{x} y-\ii\theta_{y}x}
        -b_{\rr-\ex}^{\dagger}n_\rr b_{\rr-\ey}
        \ee^{\ii\theta_{x} y+\ii\theta_{y}x}
        +\hc
    \Big),
\end{align}
where the second line results from a 180$^\circ$ rotation of the elements in the first row.

\section{More on graviton weights}
\label{app:more}

In this supplement we provide a few additional results for the graviton spectral densities recovered by using the lattice ansatz.
We expand in two fronts: simulations with intermediate fluxes $1/7$ and $1/5$, and the behavior of the operator at high energies.

\begin{figure}
    \centering
    \begin{overpic}[width=0.49\linewidth]{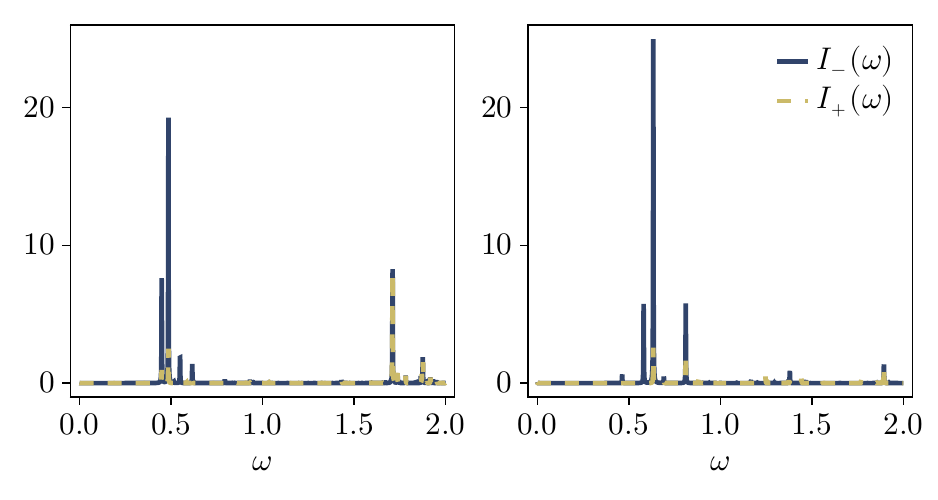}
        \put(10,44){(a)}
        \put(30,35){\footnotesize$n_\phi=1/7$}
        \put(30,30){\footnotesize$10\times7$}
        \put(60,44){(b)}
        \put(80,30){\footnotesize$n_\phi=1/5$}
        \put(80,25){\footnotesize$10\times5$}
    \end{overpic}
    \begin{overpic}[width=0.49\linewidth]{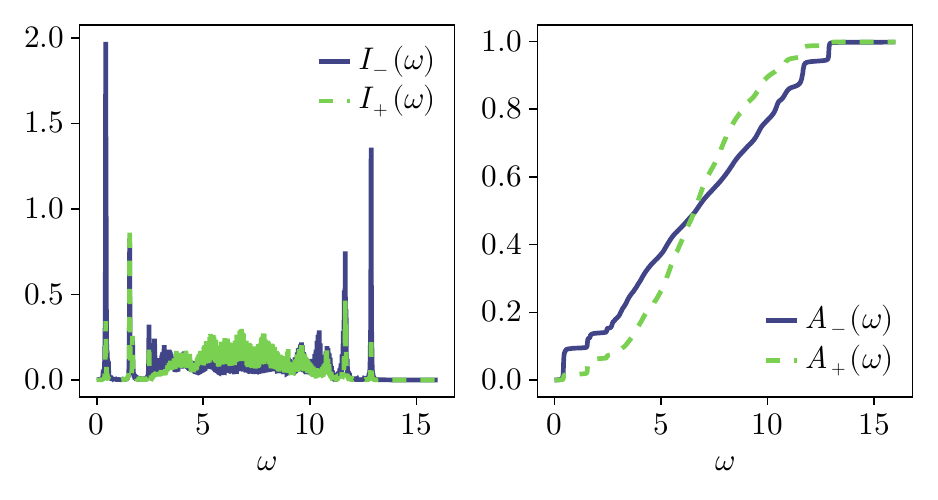}
        \put(16,44){(c)}
        \put(20,30){\footnotesize$n_\phi=1/8$}
        \put(20,25){\footnotesize$12\times8$}
        \put(62,44){(d)}
    \end{overpic}
    \caption{More on graviton weights.
    Low-energy spectral densities $I_\pm(\omega)$ for fluxes (a) $n_\phi=1/7$ and (b) $n_\phi=1/5$. 
    (c) High-energy behavior for flux $n_\phi=1/8$.
    (d) Integrated density over frequencies.
    ED results on tori. Spectral densities obtained by taking $n_\mrm{Krylov}=1000$ states. The Lorentzian broadening is set to $\gamma=0.001$ for panels (a) and (b), but is lowered to $\gamma=0.01$ for plots (c) and (d).
    }
    \label{fig:moreweights}
\end{figure}

Let us first consider odd-$q$ fluxes $n_\phi=1/q$, given by $q=7$ and $q=5$. 
We study tori of sizes given by $10\times q$, so that the number of particles is fixed to $N=5$.
The results are shown in Figs. \ref{fig:moreweights}(a) and (b).
Both simulations are compatible with a well-defined graviton mode. As pointed out in the main text, here we also observe graviton energies shift up as we increase the flux. From $\omega_\mrm{G}\approx0.48J$ at $n_\phi=1/7$ to $\omega_\mrm{G}\approx0.64J$ for flux $1/5$.
Collecting these results along with the approximate data for fluxes $1/8$ and $1/6$, and assuming a linear relation between them, we estimate the graviton energy as
$\omega_\mrm{G}\approx3.3\times n_\phi J$.
Given the definition of the pseudopotential strength $V_0=8n_\phi J$, we obtain the direct relationship
$\omega_\mrm{G}\approx0.42\times V_0$.

Next we consider the high-energy behavior of the weights $I_{\pm}(\omega)$. For it we turn again to the $n_\phi=1/8$ system of size $12\times8$ --  other fluxes are considered with MPS methods below.
In Figs. \ref{fig:moreweights}(c) we plot the graviton weights across the high-energy spectrum. This is to be contrasted with Fig. \ref{fig:weights}(a), where only the low-energy window $\omega\lesssim2$ is plotted.
Despite the broad weight in the middle of the spectrum, we still clearly observe the chiral graviton signal at low energies.
Figure \ref{fig:moreweights}(d) shows the corresponding accumulated weight $A_\pm(\omega)=\int_0^{\omega}\dd\omega'\,I_\pm(\omega')$
over the spectra.
Starting at zero frequency, we observe $A_-$ features jumps coming from the low-energy graviton peaks. The recovered weight for $I_-(\omega)$ at $\omega=4\pi Jn_\phi\approx1.6$ is about $A_-\approx0.12$. 
Leaving the low-energy realm, both $A_-$ and $A_+$ rapidly ramp until we eventually recover the full spectral densities around energies $\omega\approx13J$.

\section{Numerical procedure for spectral functions with MPS}
\label{app:numerical}

In this supplement we provide the details on the calculation of graviton spectral functions in open geometries, i.e., cylinders and rectangles, carried out with MPS techniques.

We first obtain the ground state $\ket{\psi_0}$ and its energy $E_0$ with DMRG \footnote{We checked that any combination of the two degenerate ground states does not change results.} with truncation errors around $10^{-8}$. Then we apply the graviton operator, say $O_-$, represented as an MPO to obtain $\ket{\psi_-}=\delta O_-\ket{\psi_0}$, again with a truncation error around $10^{-8}$ to keep a small bond dimension. Note that in open geometries we remove the ground state expectation value from $O_-$ to remove spectral weight at $\omega=0$, i.e., we use the operator $\delta O_-=O_- - \langle\psi_0|O_-\ket{\psi_0}$. The state is then evolved to obtain $\ket{\psi_-(t)}=\ee^{-\ii Ht}\ket{\psi_-}$ with TDVP, in particular a single-site algorithm with global subspace expansions every 5 steps, with a time step of $\delta t=0.02/J$ and a maximum bond dimension of $\chi=300$. The spectral information of the operator $\delta O_-$, i.e., the normalized graviton weight $\delta I_-(\omega)$ can then be recover by Fourier transforming the overlap $\langle\psi_-|\psi_-(t)\rangle$. In particular it is possible to show that:
\begin{align}
    \delta I_-(\omega)= \frac{1}{\langle \psi_-|\psi_-\rangle}\sum_n|\langle \psi_n|\delta O_-|\psi_0\rangle|^2 \delta(\omega-E_n+E_0)
    =\lim_{\gamma\to 0}\;\frac{1}{2\pi} \int_{-\infty}^{\infty} \mathrm{d}t\;  \frac{\langle \psi_-|\psi_-(t)\rangle }{\langle \psi_-|\psi_-\rangle}\ee^{\ii(E_0+\omega)t}\ee^{-\gamma|t|},
\end{align}
where $\gamma$ is a Lorentzian regulator for the delta function and negative times can just be obtained by complex conjugating the overlap. The normalization ensures that $\int_0^\infty \mathrm{d}\omega\, \delta I_-(\omega)=1$. In practice we evolve for a finite time $T=100/J$ and take $\gamma =5/T=0.05 J$, setting a small but finite frequency resolution. Note that not removing the ground state expectation value of $O_-$ just generate a peak at $\omega=0$ with weight $|\langle \psi_0|O_-\ket{\psi_0}|^2$ and does not change any of the finite frequency weights. MPS techniques allows us to reach larger volumes (up to $N=12$) with respect to what studied in the main text with ED (up to $N=8$), at the expenses of a finite frequency resolution of $\gamma=0.05J$.

\begin{figure}
    \centering
    \begin{overpic}[width=0.48\linewidth]{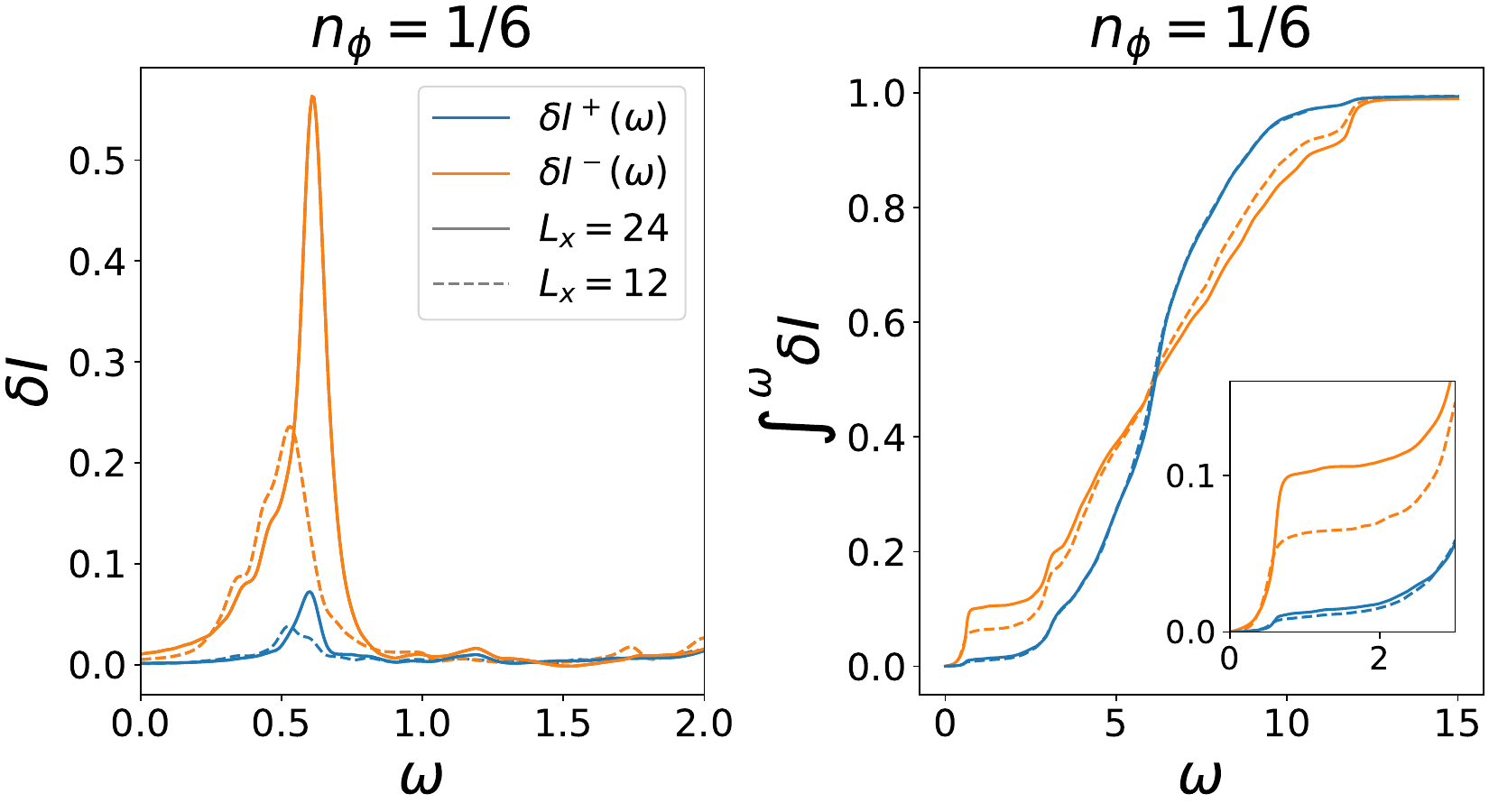}
    \put(10,45){(a)}
    \put(63,45){(b)}
\end{overpic}
\begin{overpic}[width=0.48\linewidth]{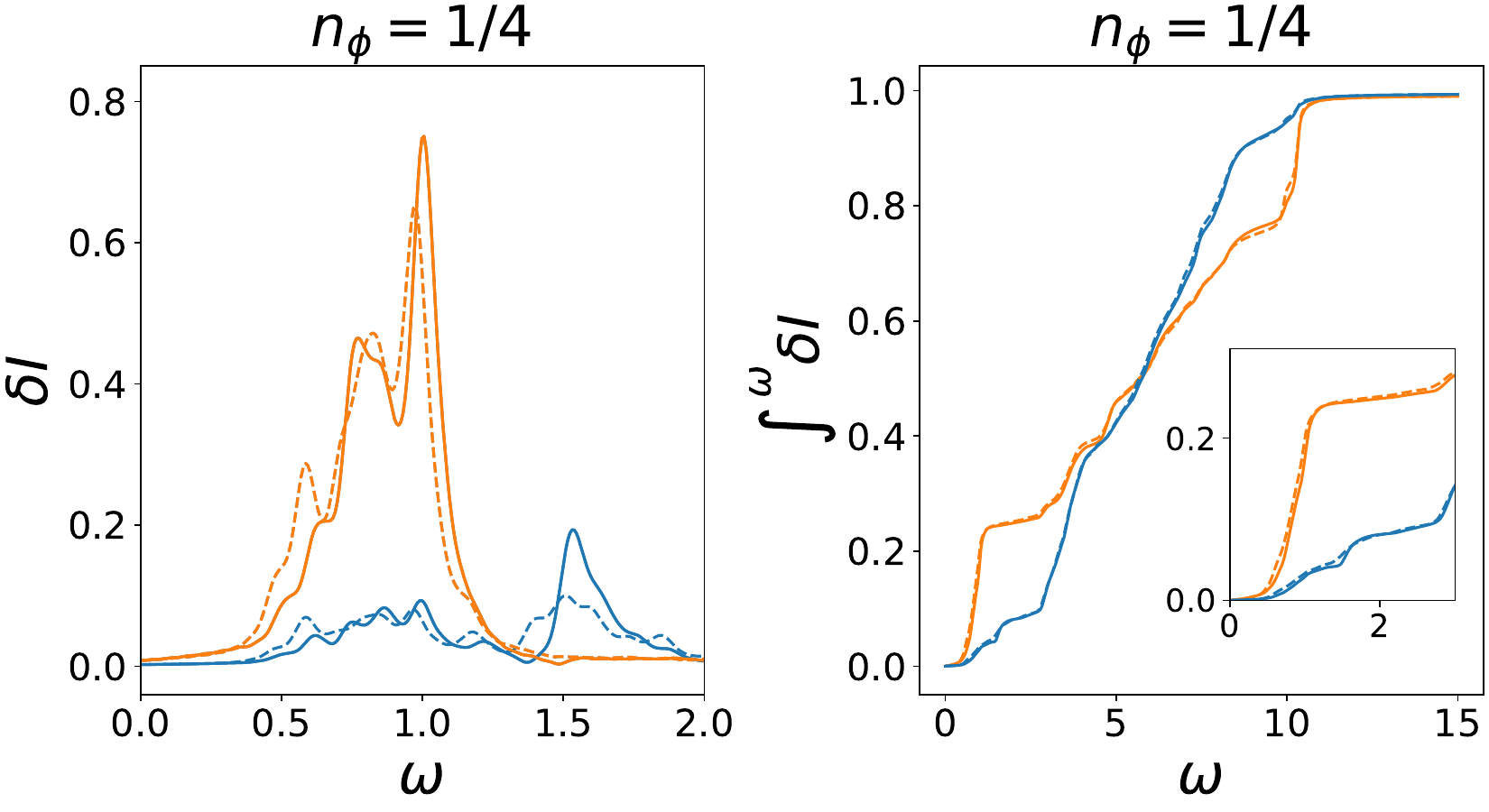}
            \put(10,45){(c)}
    \put(63,45){(d)}
    \end{overpic}
    \caption{Chiral graviton spectral function $\delta I_\pm(\omega)$ and its integral $\int_0^\omega d\omega'\delta I_\pm(\omega')$ obtained with TDVP. In (a,b) $n_\phi=1/6$ and (c,d) $n_\phi=1/4$ for cylinders of length $L_x=12,24$ (dashed,full) and circumference $L_y=1/n_\phi$. Inset in (b) and (d) show the integrated spectral function zoomed at small frequencies. }
    \label{fig:mps_weights}
\end{figure}

\section{Large volume graviton spectral functions}
\label{app:large}

In this section we provide extra results obtained with MPS techniques which corroborate the picture discussed in the main text.

In Fig.\ \ref{fig:mps_weights} we show results on (a,c) the chiral graviton spectral function $\delta I_\pm(\omega)$ and (b,d) its integrated weight $\int_0^\omega \mathrm{d}\omega' \delta I_\pm(\omega')$ for $n_\phi=1/6$ and $n_\phi=1/4$ on cylinders of circumference $L_y=1/n_\phi$ and length $L_x=12,24$. Both signals at $n_\phi=1/6$ and $n_\phi=1/4$ are strongly chiral smaller at small frequencies $\omega\lesssim2$ which roughly correspond to the cyclotron energy of the continuum limit. This is even clearer in the integrated weight which shows a plateux around $\omega \simeq 2$ dividing an effective LLL from the rest of the spectrum. The weight collected up to such frequency remains however quite small. The leading effect is likely the presence of Landau Level mixing, something which has so far always being neglected also in continuum models. Thus even in presence of a perfect continuum limit $n_\phi\to0$ the LLL projection is an approximation and the total weight of the graviton operators is not guaranteed to lie exclusively in the LLL. Finite size, both in $L_x$ and $L_y$, could also play a role in this respect. The difference in life-time of the graviton-mode between $n_\phi=1/6$ and $n_\phi=1/4$ is consistent with what discussed in the main text, i.e., at $n_\phi=1/4$ there is a strong mixing with the continuum and the weight is broad. Note that here the cylinder edges allows for new scattering channels for the graviton mode also at $n_\phi=1/6$ where indeed the peak gets sharper as $L_x$ is increased. The broad distribution of weight for the $n_\phi=1/4$ is instead less affected by the increase in system size, suggesting that the scattering mechanism is of bulk origin.

\section{Small volume graviton spectral functions}
\label{app:small}

\begin{figure}
    \centering
    \begin{overpic}
        [width=0.3\linewidth]{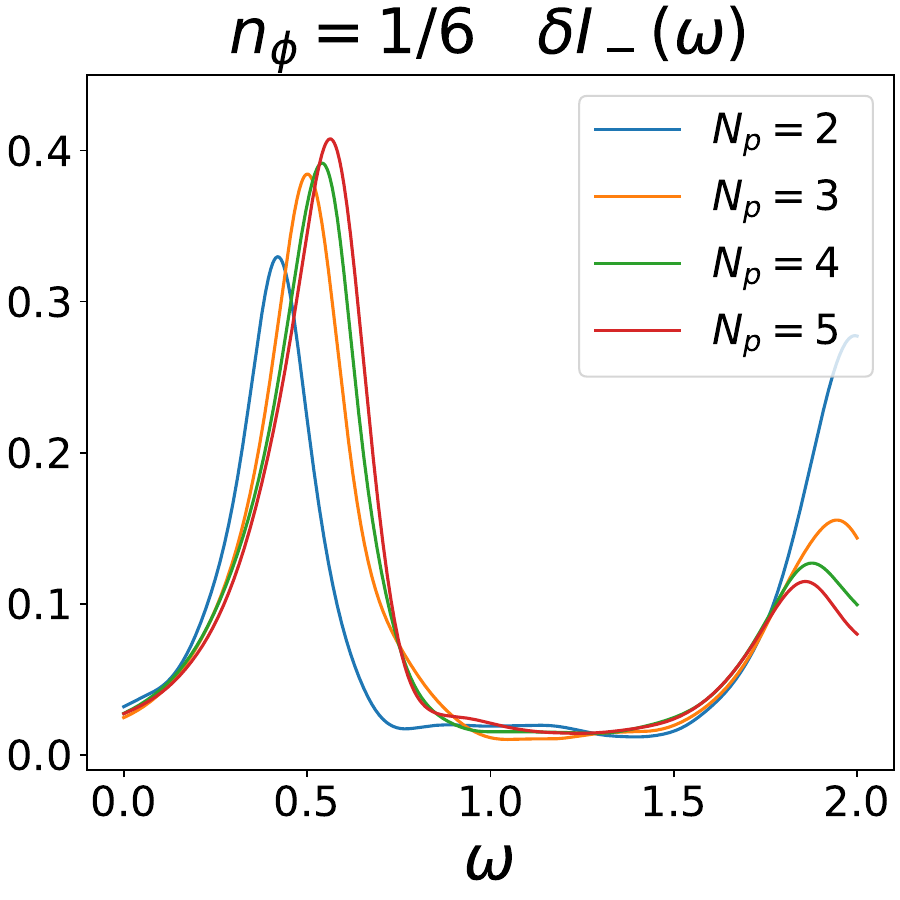}
        \put(0,0){(a)}
    \end{overpic}
        \begin{overpic}
        [width=0.3\linewidth]{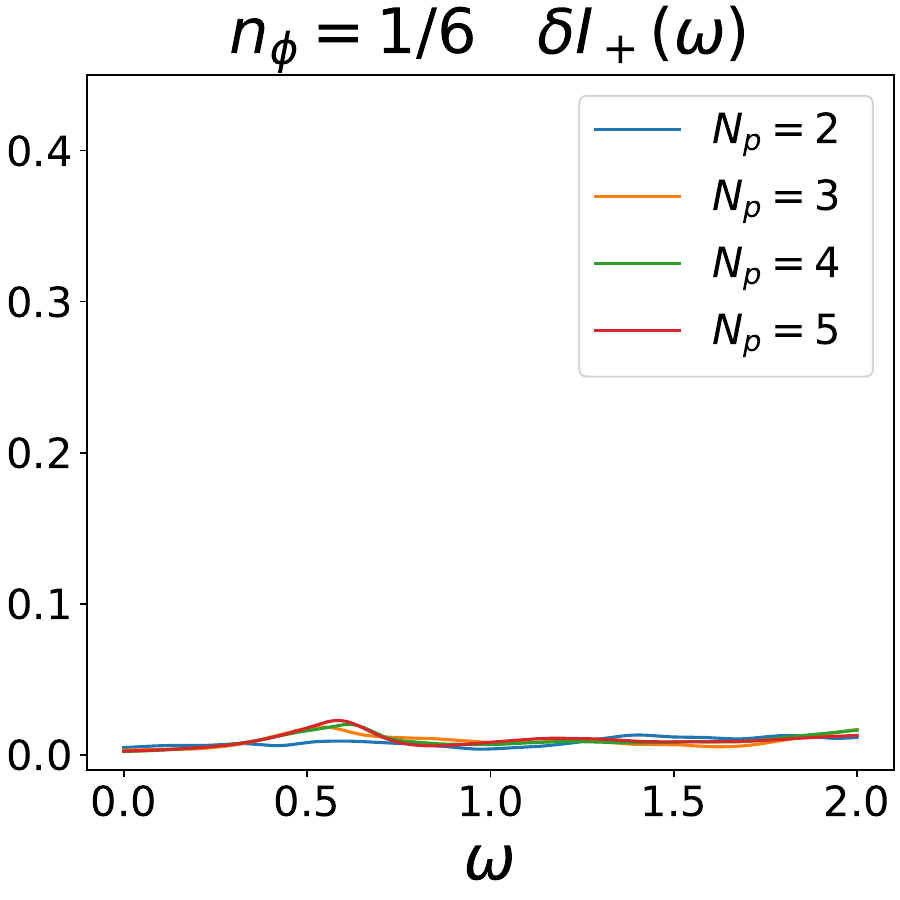}
        \put(0,0){(b)}
    \end{overpic}
    \caption{Chiral graviton spectral weights on a square geometry obtained with TDVP for small number of particles. The system sizes are $L\times L$ with $L=7,8,9,10$ respectively for $N=2,3,4,5$. The flux is fixed to $n_\phi=1/6$.}
    \label{fig:small_volume}
\end{figure}

In this section we discuss the fate of the the graviton-mode for small systems on an open square geometry. 
In Fig. \ref{fig:small_volume} we show the chiral graviton weights for systems with $N=2,3,4$, and $5$ particles. This discussion is particularly relevant  in light of recent experimental advances in realizing Laughlin states using rotating optical tweezers~\cite{lunt_prl2024_fqhfermions}, which enable high-fidelity access to few-particle systems at effectively zero temperature.

We confirm that the graviton weight exhibits a peak even for $N=2$. To better understand this, it is helpful to step back and examine the $N=2$ case in the continuum limit with contact LLL interactions, denoted by $V_0$. Two-particle states in the LLL can be expressed in terms of total and relative angular momentum $\ket{M,m}=(z_1+z_2)^M (z_1-z_2)^m \ee^{-(z_1^2+z_2^2)/4\ell_B^2}$ where $z_j=x_j+\ii y_j$ are the particle coordinates and we work in the symmetric gauge. The Laughlin state corresponds to $\ket{0,2}$. Contact interactions take the form $H=V_0\sum_M \ket{M,0}\bra{M,0}$, while the (unnormalized) chiral graviton operators are given by $\mathcal{O}_-=\sum_M \ket{M,0}\bra{M,2}$ and $\mathcal{O}_+=\sum_M \ket{M,2}\bra{M,0}$ \cite{liou2019chiral}. Together, these imply that $\mathcal{O}_+$ annihilates the Laughlin state, while $\mathcal{O}_-$ creates the excited state $\ket{0,0}$. The latter is an exact eigenstate of the contact interaction Hamiltonian, with energy $V_0$. In the many-body limit, this excitation evolves into the so called graviton mode, which appears at energy $\omega\sim1.4V_0$ for $N$ large \cite{liou2019chiral}.

It is then straightforward to interpret the few-particle results in the lattice in terms of what we know from the continuum case. Small corrections arising from the lattice lead to finite weights on the $O_+$ lattice operator. The appearance of higher LL weight around $\omega\sim 1.8$ and $2.0$ indicates that the lattice operator also excites states in higher LLs. This effect however diminishes as the number of particles is increased. Additionally, we note that the upward shift of the graviton excitation with increasing $N$ mirrors the behavior seen in the continuum case. 
We emphasize that the presence of a well-defined excitation at $N=2$ is not particularly surprising, as there is no two-magnetoroton many-body continuum for it to decay into. The key point is that, as the many-body limit is taken $N\gg 1$, this excitation shall develop into a collective mode, whose mixing with other states remains insufficient to completely wash it out. 

\section{Phase diagram on cylinders with MPS}
\label{app:phase}

In this appendix we provide extra simulations on the phase diagram at $\nu=1/2$ obtained with DMRG on cylinders of size $L_x\times L_y$ with $n_\phi=1/6$ and a commensurate $L_y=6$. The maximum bond dimension is $\chi=500$ which ensures a truncation error below $10^{-6}$, reached in the low $\eta$ regime.

\begin{figure}
    \centering
    \includegraphics[width=\linewidth]{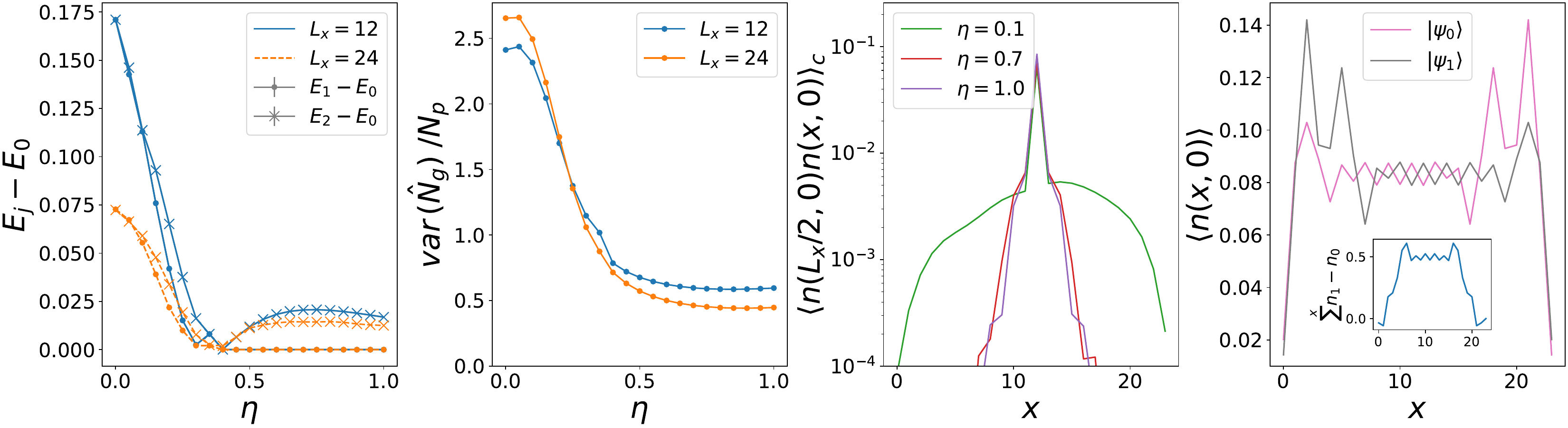}
    \caption{Phase diagram as a function anisotropy $\eta$ obtained with DMRG at $n_\phi=1/6$ for cylinders of circumference $L_y=6$. (a) First two energy gaps for two cylinder length $L_x$. (b) Graviton number $N_G=(O_++O_-)(O_-+O_+)/4$ variance divided by the number of particles $N=\nu n_\phi L_xL_y$. (c) Connected intra-chain density density correlation function for the FQH phase ($\eta=0.7$ and $\eta=1$) and close to the decoupled chain limit ($\eta=0.1$). (d) Particle density of the two ground states $\ket{\psi_0}$ and $\ket{\psi_1}$ in the topological phase $\eta=0.7$. Inset shows the integrated density difference between these two $\sum_{y,x'<x} \langle n(x',y)\rangle_1 -\langle n(x',y)\rangle_0$, highlighting the shift of a quasi-particle charge $0.5$ when passing from one ground state to the other. }
    \label{fig:mps_pd}
\end{figure}

\paragraph{Energy gaps}As shown in Fig.\ \ref{fig:mps_pd}(a), the ground state is degenerate in the FQHE phase $\eta \gtrsim 0.3$ while it is non-degenerate for the other phase, connected to the decoupled chain limit at $\eta=0$. Apart from the topological degeneracy, we see a small gap in the FQHE phase attributed to an edge state (ED on torus shows a much larger gap) which indeed does not significantly change in magnitude when $L_x$ is increased. On the contrary the gap of the decoupled chain phase decreases as roughly half once $L_x$ is doubled. This supports the interpretation of weakly coupled chains oriented on the $x$ direction where the hopping is strong \footnote{The stability of this phase and its true nature in the thermodynamic limit $L_y\to \infty$ is an interesting question beyond the scope of this work.}. 

\paragraph{Graviton number variance}In Fig.\ \ref{fig:mps_pd}(b) we show that the variance of the graviton number defined as $N_G=(O_++O_-)(O_-+O_+)/4$ also roughly indicated a very similar transition point. Indeed in the FQH phase the existence of a good graviton excitation, i.e., sharp spectral response, means that one can think of $O_\pm$ as raising and lowering operators thus leading to no variance of its number, akin to what happens for an harmonic oscillator in its ground state. Importantly we see that the residual fluctuations per particle in the FQHE phase are compatible with a finite size effect as they diminish once the system size is increased. 

\paragraph{Intra-chain correlations}In Fig.\ \ref{fig:mps_pd}(c) we also show intra-chain density correlations (at $y$ fixed) which show (i) the bulk gapped nature of the FQH phase with an exponential decay and (ii) the slower decay in the decoupled chain phase at small $\eta$ compatible with a collection of Luttinger liquids forming in the $L_x\to \infty$ limit. 

\paragraph{Density profile of FQH}Last, in Figure \ref{fig:mps_pd}(d) we show the density profile as a function of $x$ ($y$ is periodic) of the two degenerate ground states in the FQH phase at $\eta=0.7$. These differ by a quasi-particle being shifted from one end to the other of the cylinder. To highlight this we show in the inset the integrated density difference $\sum_{y,x'<x} \langle n(x',y)\rangle_1 -\langle n(x',y)\rangle_0$ between the two ground states ($0$ and $1$) which clearly shows the value $0.5$ expected for the accumulation of a charge corresponding to a quasi-particle. We further note the presence of bulk CDW oscillations compatible with a finite $L_y$ effect. Similar oscillations are indeed also present for the continuum Laughlin state when placed on thin cylinders and vanish once $L_y\gg l_B$. Interestingly we report that for $\eta=0.7$ these oscillations are actually smaller in amplitude than at $\eta=1$ (not shown), compatible with a diminished finite size effect coming from the anisotropy of the state which is realized.

\end{document}